\documentclass[twocolumn,11pt]{article}
\usepackage{bm,color}
\usepackage{graphicx}
\usepackage{amsmath,amssymb}
\usepackage{braket}
\usepackage{here}
\begin{document}
\title{\bf Theoretical studies on the spin-charge dynamics in Kondo-lattice models}
\author{Masahito MOCHIZUKI, and Rintaro ETO\\
{\it Department of Applied Physics, Waseda University}\\
{\it Okubo, Shinjuku-ku, Tokyo 169-8555} }
\date{}
\maketitle
\thispagestyle{empty}
\abstract{The Kondo-lattice model describes a typical spin-charge coupled system in which localized spins and itinerant electrons are strongly coupled via exchange interactions and exhibits a variety of long-wavelength magnetic orders originating from the nesting of Fermi surfaces. Recently, several magnetic materials that realize this model have been discovered experimentally, and they have turned out to exhibit rich topological magnetic phases including skyrmion crystals and hedgehog lattices. Our recent theoretical studies based on the large-scale spin-dynamics simulations have revealed several interesting nonequilibrium phenomena and excitation dynamics in the Kondo-lattice model, e.g., dynamical magnetic topology switching and peculiar spin-wave modes of the skyrmion crystals and the hedgehog lattices under irradiation with electromagnetic waves such as microwaves and light~\cite{Eto2021,Eto2022,Eto2024}. These achievements are expected to open a new research field to explore novel nonequilibrium topological phenomena and related material functions of the spin-charge coupled magnets.}

\section{Introduction}
Since the discovery of magnetic skyrmions in chiral magnets in 2009~\cite{Muhlbauer2009,XZYu2010}, topological spin textures such as skyrmions, merons, biskyrmions, hedgehogs, and hopfions have attracted a great deal of research interest from the viewpoints of both fundamental science and technical applications~\cite{SekiBook2016,Nagaosa2013,Everschor2019,Gobel2021,Tokura2021}. A variety of physical phenomena and material functions of the topological magnetism have been intensively studied and explored, which include magnetic memory device functions based on their current-driven motion~\cite{Fert2013,Iwasaki2013a,Iwasaki2013b,Finocchio2016,SeidelBook2016}, various anomalous quantum transport phenomena induced by emergent magnetic fields through the Berry phase mechanism, unique optical and microwave device functions arising from their peculiar excitation dynamics~\cite{Mochizuki2012,Mochizuki2013,Okamura2013,Mochizuki2015a,Mochizuki2015b}, and potentially useful power generation functions originating from coupling with electric charges~\cite{Ohe13,Shimada15,Koide2019,Matsuki2023}.

Many of the topological magnetic textures with spatially modulated magnetizations are manifested and stabilized by the Dzyaloshinskii-Moriya interaction (DMI)~\cite{Dzyaloshinsky1958,Moriya1960a,Moriya1960b}, which becomes active in magnets with broken spatial inversion symmetry, e.g., chiral magnets~\cite{Muhlbauer2009,XZYu2010,XZYu2011,Seki2012a,Seki2012b}, polar magnets~\cite{Kezsmarki2015,Kurumaji2017,Kurumaji2021}, and magnetic heterojunctions~\cite{Fert2013}. The DMI has a relativistic origin and favors a helical alignment of magnetizations with a 90-degree rotation angle. This interaction competes with the ferromagnetic exchange interaction, which favors parallel alignment of magnetizations. Their competition results in the formation of topological magnetic textures with moderate pitch angles.

In such DMI magnets, the skyrmions can appear not only as isolated defects in the ferromagnetic background but also as a long-range order with a hexagonally crystallized form. However, other topological magnetic textures mostly appear as isolated defects only, and they are usually not stable energetically. In addition, in the DMI magnets, the spatial configurations of magnetizations are governed by structures of the DM vectors, which is predominantly determined by crystal symmetry~\cite{Bogdanov1989,Bogdanov1994}. As a result, the internal degrees of freedom such as helicity, chirality, and vorticity are generally frozen and cannot be changed.

Recently, in addition to the DMI magnets, the Kondo-lattice system, in which itinerant electrons and localized spins are coupled via exchange interactions, has been theoretically proposed as a new host of topological magnetism stabilized by different physical mechanisms~\cite{Ozawa2016,Hayami2017,Ozawa2017}. Specifically, various novel topological magnetic textures stabilized by the RKKY-type long-range spin-spin interactions mediated by itinerant electrons have been theoretically proposed, which include not only usual skyrmion crystals with a topological charge of $|Q|$=1 but also nontrivial skyrmion crystals with a larger topological charge of $|Q| \ge$2, meron crystals with a half-integer topological charge $|Q|$=1/2, and hedgehog lattices with periodically aligned hedgehog and antihedgehog pairs~\cite{Hayami2021R}. 

These topological magnetisms are realized as superpositions of magnetic helices with multiple propagation vectors ($\bm Q$ vectors) governed by Fermi-surface nesting. This indicates possible realizations of more diverse topological magnetic textures by tuning the Fermi-surface geometry and the $\bm Q$ vectors through variations in material parameters, external fields, and element substitutions. Indeed, recent experiments have observed three-dimensional hedgehog lattices in MnGe~\cite{Kanazawa2016}, MnSi$_{1-x}$Ge$_x$~\cite{Fujishiro2019} and SrFeO$_3$~\cite{Ishiwata2020}, triangular skyrmion crystals in Gd$_2$PdSi$_3$~\cite{Kurumaji2019}, Gd$_3$Ru$_4$Al$_{12}$~\cite{Hirschberger2019} and EuPtSi~\cite{Kaneko2019}, and square skyrmion crystals in GdRu$_2$Si$_2$~\cite{Khanh2020}.

More importantly, in the Kondo-lattice magnets, the topological magnetic textures do not require breaking of spatial inversion symmetry and the resulting DMI for their emergence and stabilization because they are realized by the exchange interactions between itinerant electrons and localized spins. As a consequence, the helicity, chirality, and vorticity of the magnetic textures remain unfrozen as internal degrees of freedom in this system, which enables us their low-energy excitations and continuous modulations. This aspect provides a unique opportunity to control and switch magnetic topology by various external stimuli. 

It is known that skyrmions and skyrmion tubes in chiral magnets exhibit interesting physical phenomena and potential device functions originating from their unique spin-wave modes in the microwave regime. We expect that the topological magnetism manifested in Kondo lattice magnets also exhibits interesting phenomena and functionalities associated with possible collective excitation modes peculiar to the Kondo-lattice system. However, nature and properties of the spin-charge dynamics have not been elucidated sufficiently. 

In this paper, we overview the following three topics associated with our recent theoretical studies in this direction.\\
\noindent
{\bf (1) Microwave-induced magnetic topology switching~\cite{Eto2021}:} It is discovered that application of circularly polarized microwave magnetic field can switch the magnetic topology among three topological magnetic phases characterized by different topological numbers ($|N_{\rm sk}|$=0, 1, and 2) in a triangular Kondo-lattice model.\\
\noindent
{\bf (2) Low-energy excitation modes and spin-charge segregation~\cite{Eto2022}:} It is discovered that a zero-field skyrmion crystal phase in a triangular Kondo-lattice model has three linearly dispersive Goldstone modes associated with spin excitations and one quadratically dispersive pseudo-Goldstone mode associated with charge excitations. Surprisingly they turn out to be perfectly decoupled in contrast to our naive expectation of strong spin-charge coupling in the Kondo-lattice model.\\
\noindent
{\bf (3) Spin-wave modes associated in the quadratic hedgehog lattices~\cite{Eto2024}:} It is discovered that quadratic hedgehog lattice phases in the cubic Kondo-lattice model have collective excitation modes associated with translational oscillation of Dirac strings connecting a hedgehog and an antihedgehog. One of the two modes turn out to vanish upon a field-induced topological phase transition where one of the two types of Dirac-string sublattices undergoes pair annihilations and disappears.

\section{Model and Method}
%%%%%%%%%%%%%%%%%%%%%%%%%%%%%%%%%%%%%%
\begin{figure}[t]
\includegraphics[scale=1.0]{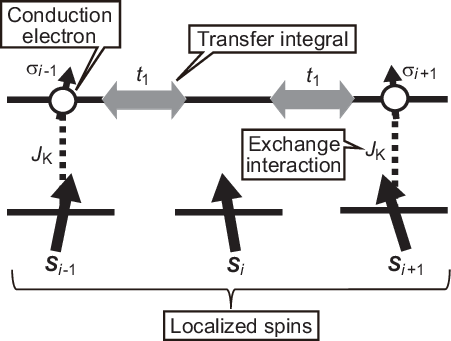}
\caption{Schematics of the Kondo-lattice model. Itinerant electrons and localized spins are coupled via the Kondo exchange coupling $J_{\rm K}$. The localized spins are not necessarily interacting directly, but indirect interactions mediated by the itinerant electrons exist, which realize long-range ordering of the localized spins.}
\label{Fig01}
\end{figure}
%%%%%%%%%%%%%%%%%%%%%%%%%%%%%%%%%%%%%%
The Hamiltonian of the Kondo-lattice model is given by,
%%%%%%%%%%%%%%%%%%%%%%%%%%%%%%%%%%%%%%
\begin{align}
\mathcal{H}_{\rm KLM}=
&\sum_{i,j,\sigma}t_{ij} \hat{c}^\dagger_{i\sigma}\hat{c}_{j\sigma} 
-\mu \sum_{i \sigma} \hat{c}^\dagger_{i\sigma}\hat{c}_{i\sigma},
\notag \\
&-\frac{J_{\rm K}}{2}\sum_{i,\sigma,\sigma'}\hat{c}^\dagger_{i\sigma}{\bm \sigma}_{\sigma\sigma'}
\hat{c}_{i\sigma'}\cdot{\bm S}_i.
\label{eq:KLM} 
\end{align}
%%%%%%%%%%%%%%%%%%%%%%%%%%%%%%%%%%%%%%
Here $\hat{c}^\dagger_{i\sigma}$ $(\hat{c}_{i\sigma})$ is a creation (annihilation) operator of an itinerant electron with spin $\sigma(=\uparrow,\downarrow)$ on site $i$. The first and second terms describe kinetic energies and a chemical potential of itinerant electrons. The third term describes the Kondo exchange coupling between the itinerant electron spins and the localized classical spins $\bm S_i$ ($|\bm S_i|=1$).

We also consider the Zeeman-coupling term,
%%%%%%%%%%%%%%%%%%%%%%%%%%%%%%%%%%%%%%
\begin{align}
\mathcal{H}_{\rm Zeeman}=-\sum_i \left[\bm H_{\rm ext} + \bm H(t) \right]\cdot{\bm S}_i,
\label{Zeeman}
\end{align}
%%%%%%%%%%%%%%%%%%%%%%%%%%%%%%%%%%%%%%
where $\bm H_{\rm ext}=(0,0,H_z)$ is a static magnetic field, and $\bm H(t)$ is a time-dependent magnetic field acting on the localized spins. The total Hamiltonian is given by $\mathcal{H}=\mathcal{H}_{\rm KLM} + \mathcal{H}_{\rm Zeeman}$. The coupling between the magnetic fields and the itinerant electrons is neglected, since we have confirmed that it does not affect the results even quantitatively in the microwave and sub-terahertz frequency regimes.

Time evolutions of the localized spins are numerically simulated using the Landau-Lifshitz-Gilbert (LLG) equation,
%%%%%%%%%%%%%%%%%%%%%%%%%%%%%%%%%%%%%%
\begin{align}
\frac{d{\bm S}_i}{dt} = -{\bm S}_i\times{\bm H}^{\rm eff}_i + \frac{\alpha_{\rm G}}{S}{\bm S}_i\times\frac{d{\bm S}_i}{dt}.
\label{eq:LLG}
\end{align}
%%%%%%%%%%%%%%%%%%%%%%%%%%%%%%%%%%%%%%
The equation is composed of two contributions. The first term describes precession of localized spins around the effective local magnetic field $\bm H^{\rm eff}_i$. The second term phenomenologically describes a dissipation effect where $\alpha_{\rm G}$ is the dimensionless Gilbert-damping coefficient. The effective field ${\bm H}^{\rm eff}_i$ is calculated by,
%%%%%%%%%%%%%%%%%%%%%%%%%%%%%%%%%%%%%%
\begin{align}
{\bm H}^{\rm eff}_i = -\frac{\partial\Omega}{\partial{\bm S}_i} + {\bm H}_{\rm ext} + {\bm H}(t).
\end{align}
%%%%%%%%%%%%%%%%%%%%%%%%%%%%%%%%%%%%%%
The thermodynamic potential $\Omega$ for $\mathcal{H}_{\rm KLM}$ is given by,
%%%%%%%%%%%%%%%%%%%%%%%%%%%%%%%%%%%%%%
\begin{align}
\Omega = \int \rho(\varepsilon,\{{\bm S}_i\}) \; f(\varepsilon-\mu)\;d\varepsilon, 
\end{align}
%%%%%%%%%%%%%%%%%%%%%%%%%%%%%%%%%%%%%%
with
%%%%%%%%%%%%%%%%%%%%%%%%%%%%
\begin{align}
&\rho(\varepsilon,\{\bm S_i\})
=\frac{1}{2N}\sum_{\nu=1}^{2N}\delta(\varepsilon-\varepsilon_\nu(\{\bm S_i\})),
\\
&f(\varepsilon-\mu)=-\frac{1}{\beta}\ln \left[1+e^{-\beta(\varepsilon-\mu)} \right],
\end{align}
%%%%%%%%%%%%%%%%%%%%%%%%%%%%%%%%%%%%%%
where $\rho(\varepsilon,\{{\bm S}_i\})$ is the density of states of the itinerant electrons for a given set of the localized spins $\{{\bm S}_i\}$, and $f(\varepsilon-\mu)$ is the free energy density of the system. We use the kernel polynomial method based on the Chebyshev polynomial expansion to calculate $\Omega$ and the automatic differentiation technique to calculate its magnetization-derivative $\partial\Omega/\partial{\bm S}_i$~\cite{Weisse2006,Barros2013}. We solve the LLG equation using the fourth-order Runge-Kutta method to obtain the spatiotemporal profiles of the dynamics of localized spins. This methodology assumes that in the low-frequency regimes of gigahertz or sub-terahertz considered here, the itinerant electrons smoothly follow the localized-spin dynamics and are immediately relaxed to a ground state for a given localized-spin configuration at each moment.

All the numerical simulations discussed below are performed at zero temperature with no thermal fluctuations in order to investigate pure effects of microwave/light irradiation for dynamical phenomena in the Kondo-lattice models. For more details of the simulations, see references indicated in each section.

\section{Microwave-Induced Switching of Magnetic Topology}
%%%%%%%%%%%%%%%%%%%%%%%%%%%%%%%%%%%%%%
\begin{figure}[h]
\includegraphics[scale=1.0]{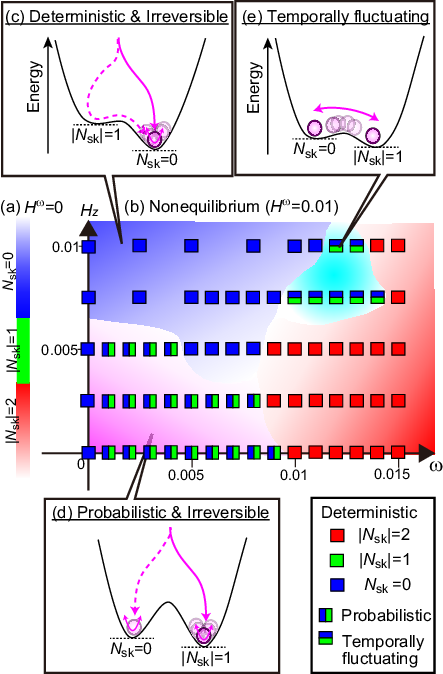}
\caption{(a) Equilibrium phase diagram of the triangular Kondo-lattice model in a static magnetic field $H_{\rm ext}=(0,0,H_z)$ without microwave  ($H^\omega$=0).  (b) Nonequilibrium phase diagram in plane of microwave frequency $\omega$ and $H_z$ under irradiation with circularly polarized microwave field. The microwave amplitude is fixed at $H^\omega$=0.01. (c)-(e) Schematic energy landscapes for different switching behaviors.}
\label{Fig02}
\end{figure}
%%%%%%%%%%%%%%%%%%%%%%%%%%%%%%%%%%%%%%
%%%%%%%%%%%%%%%%%%%%%%%%%%%%%%%%%%%%%%
\begin{figure}[t]
\includegraphics[scale=1.0]{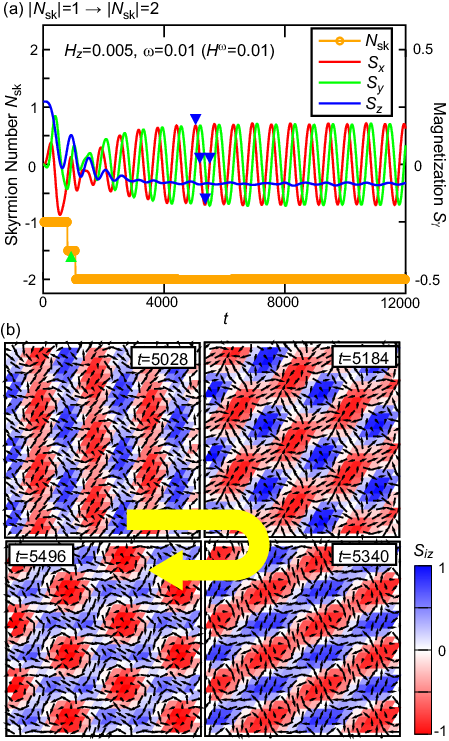}
\caption{Time profiles of the net spin $\bm S=(S_x,S_y,S_z)$ and the skyrmion number $N_{\rm sk}$ during a process of the microwave-induced topology switching from $N_{\rm sk}=-1$ to $N_{\rm sk}=-2$ in the triangular Kondo-lattice model at $H_z$=0.005 irradiated with circularly polarized microwave field with a frequency of $\omega$=0.01.}
\label{Fig03}
\end{figure}
%%%%%%%%%%%%%%%%%%%%%%%%%%%%%%%%%%%%%%
First, we discuss microwave-induced dynamical switching of magnetic topology in a triangular Kondo-lattice model~\cite{Eto2021}. We adopt $t_1=-1$ for the nearest-neighbor hopping, $t_3=0.85$ for the third-neighbor hopping, $J_{\rm K}=1$ for the Kondo exchange coupling, and $\mu=-3.5$ for the chemical potential. With this parameter set, the model is known to exhibit two types of skyrmion crystals with different skyrmion numbers as superpositions of three magnetic helices~\cite{Ozawa2017}.

The noncoplanar spin structures generally have finite local scalar spin chirality $\chi_i$, which is defined by a solid angle spanned by three neighboring spins. On the triangular lattice, it is defined by
%%%%%%%%%%%%%%%%%%%%%%%%%%%%
\begin{align}
\chi_i=\bm S_i \cdot (\bm S_{i+\hat{\bm a}} \times \bm S_{i+\hat{\bm b}})/4\pi,
\end{align}
%%%%%%%%%%%%%%%%%%%%%%%%%%%%
where $\hat{\bm a}$ and $\hat{\bm b}$ are primitive translation vectors of the triangular lattice. This quantity is related with a topological invariant called the skyrmion number $N_{\rm sk}$ as,
%%%%%%%%%%%%%%%%%%%%%%%%%%%%
\begin{align}
N_{\rm sk}=N_{\rm muc} \sum_{i=1}^N \chi_i/N,
\end{align}
%%%%%%%%%%%%%%%%%%%%%%%%%%%%
where $N_{\rm muc}$ is the number of sites in a magnetic unit cell. 

Figure~\ref{Fig02}(a) shows an equilibrium phase diagram as a function of $H_z$, in which three magnetic phases, i.e., a skyrmion crystal phase with $|N_{\rm sk}|$=2,  another skyrmion crystal phase with $|N_{\rm sk}|$=1,  and a nontopological phase with $|N_{\rm sk}|$=0, successively appear with increasing $H_z$. Importantly, several internal degrees of freedom, e.g., helicity, chirality, and vorticity, remain unfrozen in this system because of the spatial inversion symmetry, and the spin structures in these phases have infinite degeneracy. This is in striking contrast to the skyrmion crystals in chiral magnets stabilized by DMI.

The continuous degeneracy due to these internal degrees of freedom enables us to realize the transformation of spin structures by application of external stimuli. To demonstrate it, we have simulated spatiotemporal dynamics of localized spins induced by an applied microwave field ${\bm H}(t) = H^\omega\beta(t)(\cos\omega t, \sin\omega t, 0)$ with a time-dependent factor $\beta(t)$ introduced for a gradual increase of the microwave amplitude. 

The numerical simulations have indeed revealed that spin excitations induced by irradiation with circularly polarized microwave can switch the magnetic topology, e.g., from the skyrmion crystal with $|N_{\rm sk}|$=1 to another skyrmion crystal with $|N_{\rm sk}|$=2 or to a nontopological magnetic state with $|N_{\rm sk}|$=0 depending on the microwave frequency. The induced nonequilibrium spin phases are summarized in the nonequilibrium phase diagram in plane of $\omega$ and $H_z$ when the microwave amplitude is fixed at $H^\omega$=0.01 [Fig.~\ref{Fig02}(b)].

The magnetic topology switching shows various behaviors, that is, deterministic irreversible switching, probabilistic irreversible switching, and temporally random fluctuation depending on the microwave frequency and the strength of external magnetic field, variety of which is attributable to different energy landscapes in the dynamical regime. 

We show a few examples of the observed topology-switching phenomena below. In Fig.~\ref{Fig03}, we present simulated time profiles of the net spin $\bm S \equiv (1/N)\sum_i \bm S_i$ and the skyrmion number $N_{\rm sk}$ in a system irradiated with circularly polarized microwave field with a frequency of $\omega$=0.01 when $H_z=0.005$. Starting with the skyrmion crystal with $N_{\rm sk}=-1$ as a ground state at $H_z=0.005$, we observe a microwave-induced switching of the magnetic topology from $N_{\rm sk}=-1$ to $N_{\rm sk}=-2$. Here the skyrmion crystal with $N_{\rm sk}=-2$ appears as a nonequilibrium steady state where the total spin oscillates in a steady manner. Interestingly, a transient magnetic state with a half-integer skyrmion number ($N_{\rm sk}=-1.5$) is observed during the switching process. 

%%%%%%%%%%%%%%%%%%%%%%%%%%%%%%%%%%%%%%
\begin{figure}[t]
\centering
\includegraphics[scale=1.0]{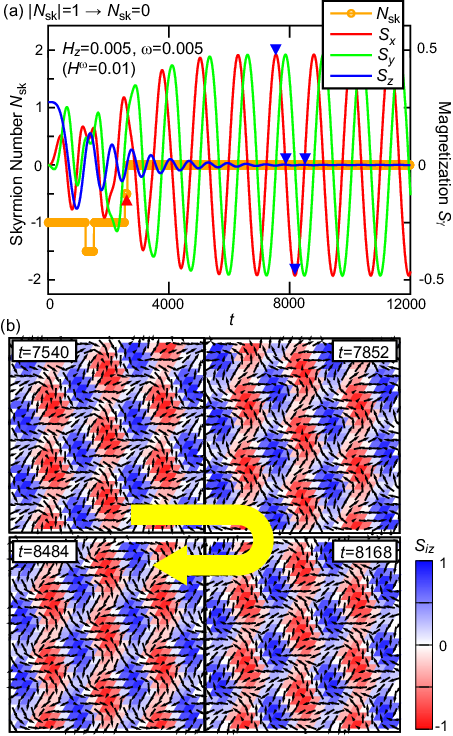}
\caption{Time profiles of the net spin $\bm S$ and the skyrmion number $N_{\rm sk}$ during a process of the microwave-induced topology switching from $N_{\rm sk}=-1$ to $N_{\rm sk}$=0 in the triangular Kondo-lattice model at $H_z$=0.005 irradiated with circularly polarized microwave with a relatively slower frequency of $\omega$=0.005.}
\label{Fig04}
\end{figure}
%%%%%%%%%%%%%%%%%%%%%%%%%%%%%%%%%%%%%%
On the other hand, Fig.~\ref{Fig04} shows simulated time profiles of the total spin $\bm S$ and the skyrmion number $N_{\rm sk}$ when the system is irradiated with a relatively slower circularly polarized microwave field with $\omega$=0.005. Starting with the a skyrmion crystal state with $N_{\rm sk}=-1$ as a ground state at $H_z=0.005$,  the nontopological state with $N_{\rm sk}=0$ appears as a nonequilibrium steady state with steady oscillation of $\bm S$. Again, a nonequilibrium transient state with a half-integer skyrmion number is observed during the switching process.

\section{Low-Energy Spin and Charge Excitations in Skyrmion Crystals}
Next we discuss our theoretically study on the spin and charge excitations of skyrmion crystals in centrosymmetric Kondo-lattice magnets~\cite{Eto2022}. As we have discussed, long-range spin interactions mediated by itinerant electrons stabilize various skyrmion-crystal phases as different patterns of superpositions of multiple spin helices whose propagation vectors are governed by the Fermi-surface nesting. Hence, we naively expect that nature of strong spin-charge coupling should appear in their low-energy excitations. However, it is not always the case in reality as argued in this section.

%%%%%%%%%%%%%%%%%%%%%%%%%%%%
\begin{figure}[t]
\includegraphics[scale=0.5]{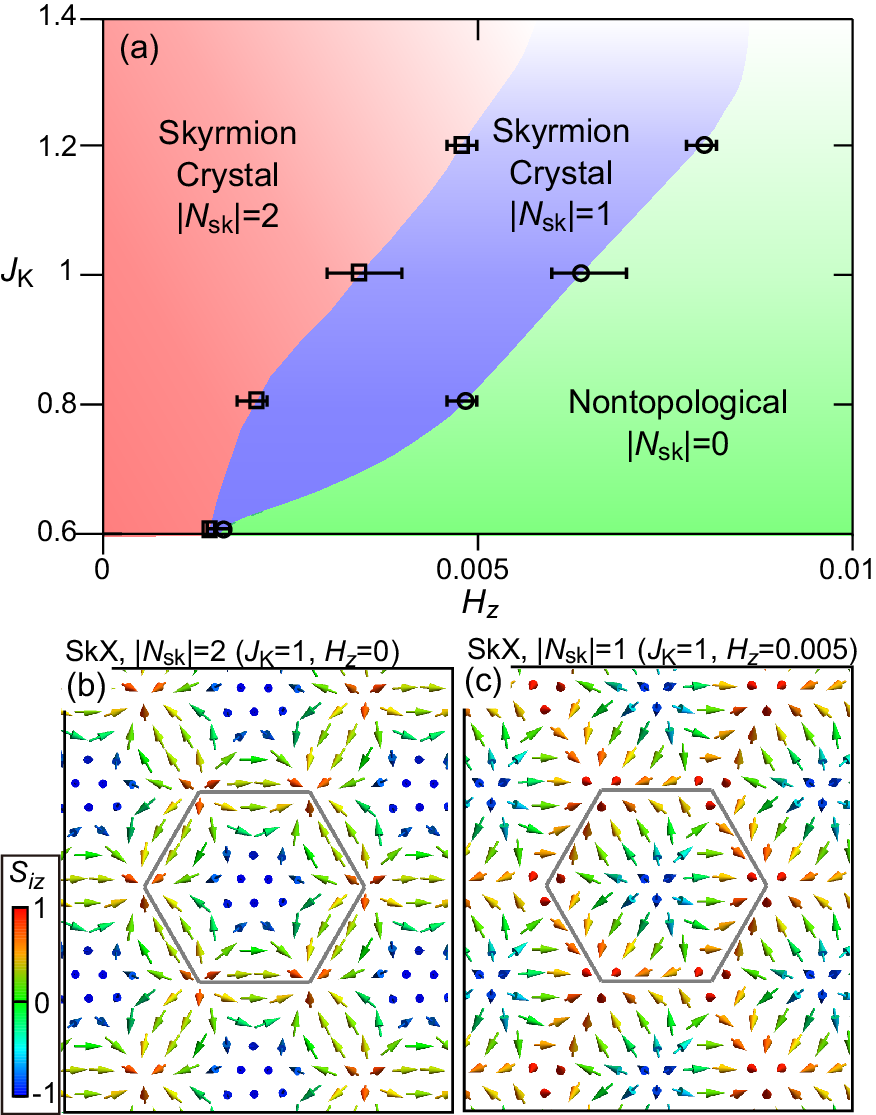}
\caption{(a) Ground-state phase diagram in plane of $H_z$ and $J_{\rm K}$ for the triangular Kondo-lattice model. (b),~(c) Spatial spin configurations $\{\bm S_i\}$ for two kinds of skyrmion crystal states with (b) $|N_{\rm sk}|$=2 and (c) $|N_{\rm sk}|$=1. For the parameter values, see text.}
\label{Fig05}
\end{figure}
%%%%%%%%%%%%%%%%%%%%%%%%%%%%
We again employ the Kondo-lattice model on a triangular lattice given by Eq.~(\ref{eq:KLM}). Figure~\ref{Fig05}(a) shows the ground-state phase diagarm in plane of $H_z$ and $J_{\rm K}$ when $t_1=-1$, $t_3=0.85$ and $\mu=-3.5$. A skyrmion crystal phase with $|N_{\rm sk}|$=2 [Fig.~\ref{Fig05}(b)], another skyrmion crystal phase with $|N_{\rm sk}|$=1  [Fig.~\ref{Fig05}(c)], and a nontopological phase with $|N_{\rm sk}|$=0 appear in this order as $H_z$ increases in the broad range of the Kondo exchange coupling $J_{\rm K}$. We set $J_{\rm K}=1 (\sim |t_1|)$, for which the system is in the weak-coupling regime and, thus, is metallic with no finite exchange gap.

To study the spin and charge excitation spectra in each phase, we simulate spatiotemporal dynamics of the localized spins $\bm S_i(t)$ after locally applying a short magnetic-field pulse using the Landau-Lifshitz equation. The equation is obtained by setting $\alpha_{\rm G}=0$ in Eq.~(\ref{eq:LLG}). The effective local magnetic field ${\bm H}^{\rm eff}_i$ is calculated again using the kernel polynomial method combined with the automatic differentiation technique~\cite{Weisse2006,Barros2013}. The wavefunction of itinerant electrons $\ket{\Psi(t)}$ at each moment $t$ is composed of eigenstates $\ket{\psi_\nu(t)}$ of the Hamiltonian $\mathcal{H}_{\rm KLM}$.

Dynamical spin and charge structure factors $S(\bm q,\omega)$ and $N(\bm q,\omega)$ are calculated from the simulated time profiles of $\bm S_i(t)$ and $\ket{\Psi(t)}$ as,
%%%%%%%%%%%%%%%%%%%%%%%%%%%%
\begin{eqnarray}
S(\bm q,\omega) \propto \sum_{i,j=1}^N e^{-i\bm q \cdot (\bm r_i-\bm r_j)}
\sum_{n=1}^{N_t} e^{i \omega t_n} \braket{\bm S_i(t) \cdot \bm S_j(0)},
\notag\\
N(\bm q,\omega) \propto \sum_{i,j=1}^N e^{-i\bm q \cdot (\bm r_i-\bm r_j)}
\sum_{n=1}^{N_t} e^{i \omega t_n} \braket{n_i(t) \cdot n_j(0)},
\notag
\end{eqnarray}
%%%%%%%%%%%%%%%%%%%%%%%%%%%%
where $n_i(t)=\sum_\sigma\bra{\Psi(t)}\hat{c}^\dagger_{i\sigma}\hat{c}_{i\sigma}\ket{\Psi(t)}$ is the expectation value of the electron number on site $i$ at time $t$. We also calculate dispersion relations of the excitation spectra by using the linear spin-wave theory.

%%%%%%%%%%%%%%%%%%%%%%%%%%%%
\begin{figure}[t]
\includegraphics[scale=0.5]{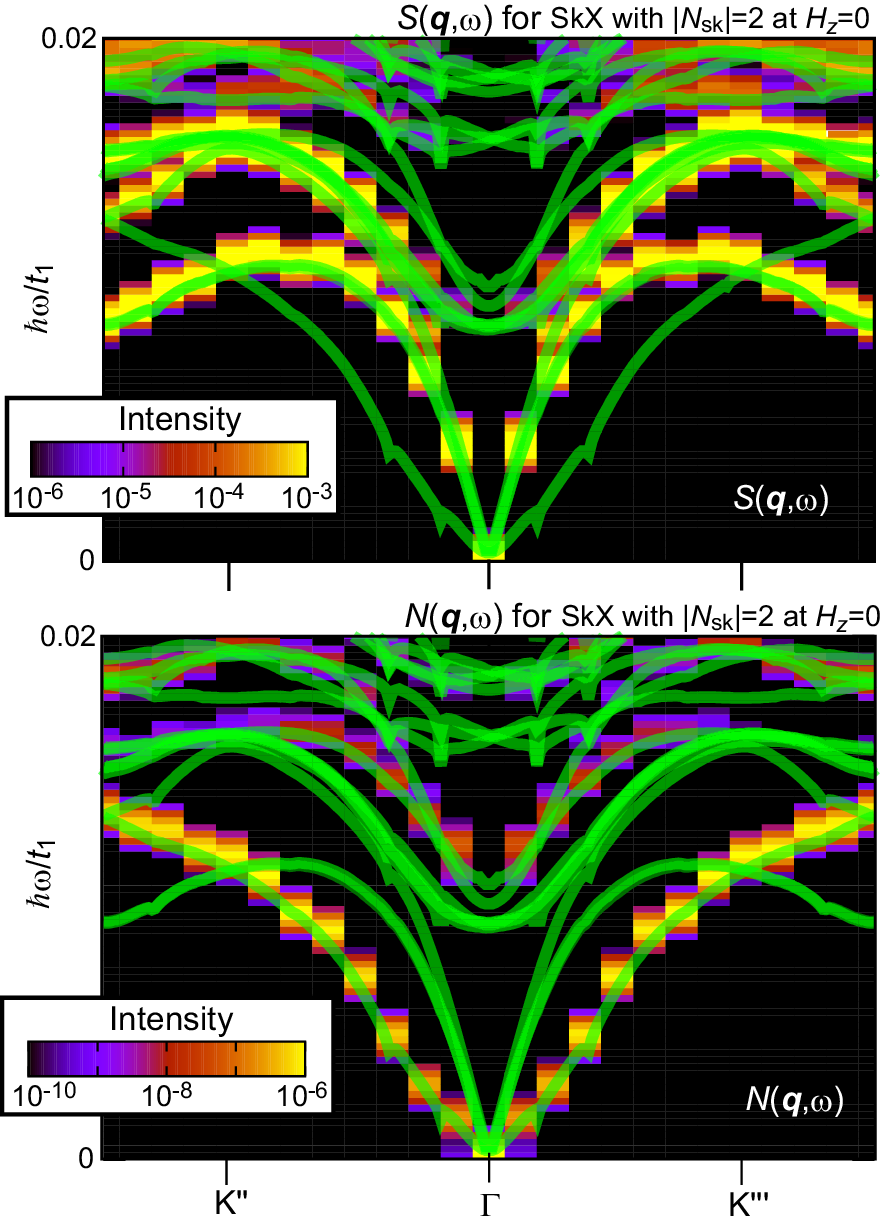}
\caption{Dynamical spin and charge structure factors, $S(\bm q,\omega)$  and $N(\bm q,\omega)$, near the $\Gamma$ point for the skyrmion crystal with $|N_{\rm sk}|=2$ at $H_z=0$. The spectra calculated by the spin dynamics simulations are presented by square dots, while the dispersion relations calculated by the linear spin-wave theory are presented by thick lines.}
\label{Fig06}
\end{figure}
%%%%%%%%%%%%%%%%%%%%%%%%%%%%
Figure~\ref{Fig06} shows calculated $S(\bm q,\omega)$ and $N(\bm q,\omega)$ for the skyrmion crystal with $|N_{\rm sk}|=2$ at $H_z=0$. Noticeably, the spin-excitation spectrum $S(\bm q,\omega)$ has large intensities on the three linearly dispersive gapless modes (Goldstone modes), whereas no intensity is observed on the quadratically dispersive mode. These three Goldstone modes are associated with three generators of the SO(3) group in the spin space and originates from spontaneous breaking of the SO(3) symmetry. 

On the other hand, the charge-excitation spectrum $N({\bm q},\omega)$ has a large intensity on the quadratic mode, whereas no intensity is observed on the three linearly dispersive Goldstone modes. This quadratic mode is associated with translational symmetry. Because of the discreteness of the lattice, this mode has a finite gap but it is negligibly small. This nearly gapless mode (pseudo-Goldstone mode) originates from spontaneous breaking of the pseudo continuous translational symmetry.

Surprisingly, it is found that the spin and charge excitations appear in different modes, which are separated and totally independent of each other in contrast to our naive expectation of the strong spin-charge coupling in the Kondo-lattice model. This spin-charge segregation in the low-energy excitations is attributable to a fact that the global SO(3) symmetry of the spin rotation and the global translation symmetry of the skyrmion-crystal spin configuration are independent and thus are decoupled. This fact may enable us selective activations of spin and charge degrees of freedom in this system as a unique property of spin textures in the centrosymmetric Kondo-lattice magnets.

However, the spin and charge degrees of freedom are not always decoupled in excitations of the Kondo-lattice magnets. In fact, the spin-charge coupling is manifested in the excitation of scalar spin chirality. The dynamical structure factor of the scalar spin chirality $C_\chi(\bm q,\omega)$ has dominant spectral weight on the quadratic mode (not shown) similar to the charge structure factor $N(\bm q,\omega)$. The dynamics of noncollinear spin texture with nonzero scalar spin chirality can induce the charge dynamics through generating emergent electromagnetic fields acting on the itinerant electrons via the Berry phase mechanism. The coincidence of charge and chirality excitation channels comes from this coupling.

%%%%%%%%%%%%%%%%%%%%%%%%%%%%
\begin{figure}[t]
\includegraphics[scale=0.5]{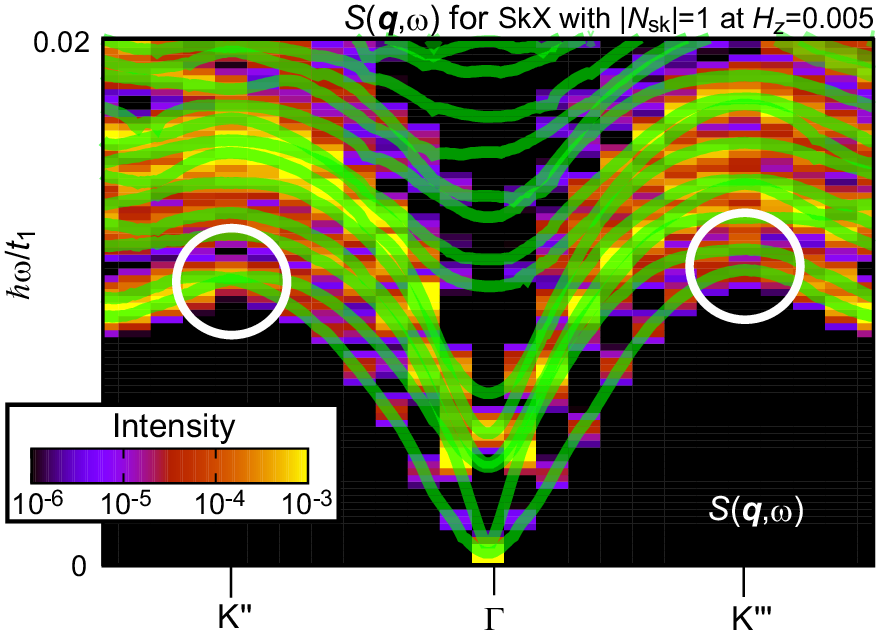}
\caption{Dynamical spin structure factor $S(\bm q,\omega)$ near the $\Gamma$ point for the skyrmion crystal with $|N_{\rm sk}|=1$ at $H_z=0.005$. Nonreciprocal nature of the spin-wave propagation appears in the dispersion relations at the momenta indicated by circles.}
\label{Fig07}
\end{figure}
%%%%%%%%%%%%%%%%%%%%%%%%%%%%
Figure~\ref{Fig07} shows the calculated dynamical spin structure factor $S(\bm q,\omega)$ for another skyrmion crystal phase with $|N_{\rm sk}|=1$ in a finite magnetic field of $H_z=0.005$. The application of magnetic field reduces the symmetry in the spin space from SO(3) to U(1). As a result, two of the three linearly dispersive Goldstone modes become gapped, and only one mode remains to be gapless. Accordingly, only one linearly dispersive Goldstone mode is observed in the spectra. On the contrary, the quadratic pseudo-Goldstone mode survives even under the application of magnetic field because the pseudo continuous translational symmetry remains even in the magnetic field. It is also found that the applied magnetic field causes mixing of the spin and charge excitations so that the spin-charge segregation observed in the $|N_{\rm sk}|$=2 phase is obscure in the $|N_{\rm sk}|$=1 phase.

The nonreciprocal nature of the spin-wave excitations, i.e., $S(\bm q,\omega) \ne S(-\bm q,\omega)$, in the skyrmion crystal phase with $|N_{\rm sk}|=1$ is worth noting as a striking feature of this system, which can be seen in comparison between the spin-wave dispersions indicated by two circles in Fig.~\ref{Fig07}. This nonreciprocity comes from symmetry of the electronic structure governed by symmetry of the skyrmion-crystal spin configuration in the magnetic field. Indeed, the Fermi surface in this phase does not have the $C_2$ point-group symmetry, and thus the symmetry of the system is reduced to $C_{3h}$ from original $D_{6h}$ for the triangular lattice. This nonreciprocal nature might be detected experimentally by e.g., neutron-scattering experiments, angle-resolved photoemission spectroscopies, and several transport measurements.

\section{Collective Modes in Quadratic Hedgehog Lattices}
%%%%%%%%%%%%%%%%%%%%%%%%%%%%%%%%%%%%%%
\begin{figure}[t]
\includegraphics[scale=0.5]{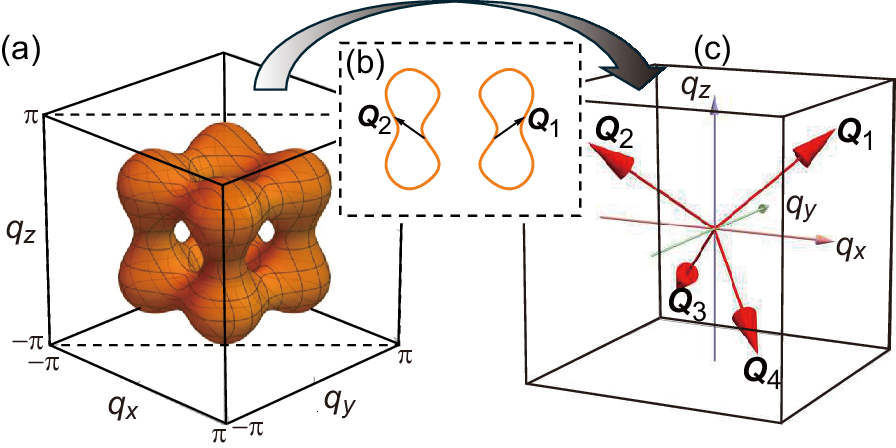}
\caption{Fermi surface for the kinetic term in Eq.~(\ref{eq:KLM}). (b) Cross section of the Fermi surface with nesting vectors $\bm Q_1$ and $\bm Q_2$. (c) Four propagation vectors $\{\bm Q_\nu\}$ of magnetic helices constituting the quadratic hedgehog-lattice (4$Q$-HL) states.}
\label{Fig08}
\end{figure}
%%%%%%%%%%%%%%%%%%%%%%%%%%%%%%%%%%%%%%
Finally, we discuss a theoretical study on the spin-wave modes of magnetic hedgehog lattices in the Kondo-lattice model in three dimensions~\cite{Eto2024}. The monopole and antimonopole are conceptual particles which were originally proposed by Paul Dirac in 1931 as isolated positive and negative magnetic charges~\cite{Dirac1931}. The magnetic hedgehog and antihedgehog are spin textures in magnets, which can be regarded as emergent monopole and antimonopole as they behave as source and sink of emergent magnetic fields acting on itinerant electrons through the Berry-phase mechanism.

The magnetic hedgehog lattices are three-dimensional magnetic structures in which the hedgehogs and antihedgehogs are aligned in a periodic manner, which have been recently discovered in several itinerant magnets, e.g., MnGe~\cite{Kanazawa2016}, MnSi$_{1-x}$Ge$_x$~\cite{Fujishiro2019}, and SrFeO$_3$~\cite{Ishiwata2020}. The hedgehog lattice in MnGe is described by a superposition of three spin helices with cubic propagation vectors and, thus, is referred to as the triple-$Q$ hedgehog lattice (3$Q$-HL). On the other hand, the hedgehog lattices in MnSi$_{1-x}$Ge$_x$ and SrFeO$_3$ are described by superpositions of four spin helices with tetrahedral propagation vectors, which is, thus, referred to as the quadruple-$Q$ hedgehog lattice (4$Q$-HL).

There are several kinds of 4$Q$-HL states with different number of hedgehogs and antihedgehogs in the magnetic unit cell. Possible transformations among them by application of magnetic field have been experimentally proposed. This proposal provides us with an opportunity to study the tunability of material properties via the field-induced topological phase transitions of the hedgehog lattices in MnSi$_{1-x}$Ge$_x$ and SrFeO$_3$. Under these circumstances, clarification of the collective spin-charge dynamics of the hedgehog lattices is a subject of essential importance because they are deeply related with responses to electromagnetic waves (e.g., microwaves and light) as sources of rich magnetic/optical device functions, dynamical topological phase transitions, and optical control of topology-related material properties. 

To study collective excitation modes in the 4$Q$-HL states in itinerant chiral magnets MnSi$_{1-x}$Ge$_x$, we start with the Kondo-lattice model on a chiral cubic lattice. The following two terms are newly added to the original Hamiltonian in Eq.~(\ref{eq:KLM}), 
%%%%%%%%%%%%%%%%%%%%%%%%%%%%%%%%%%%%%%%%%%%%%%%
\begin{align}
&\mathcal{H}_{\rm DMI}=
-D\sum_{<i,j>}\bm e_{ij} \cdot (\bm S_i \times \bm S_j),
\\
&\mathcal{H}_{\rm AFM}=
\sum_{<i,j>}J_{\rm AFM}\bm S_i \cdot \bm S_j.
\end{align}
%%%%%%%%%%%%%%%%%%%%%%%%%%%%%%%%%%%%%%%%%%%%%%%
The term $\mathcal{H}_{\rm DMI}$ describes the DMI, while the term $\mathcal{H}_{\rm AFM}$ describes antiferromagnetic exchange interaction, both of which work between the nearest-neighbor localized spins. Here $\bm e_{ij}$ denotes the normalized bond-directional vector from site $i$ to site $j$. We adopt $J_{\rm K}$=0.8, $J_{\rm AFM}$=0.0008 and examine both cases with and without DMI (i.e., $D=0$ and $D=0.0002$). Note that the DMI generally originates from the spin-orbit coupling, but its effects on the itinerant electrons are neglected here for simplicity, which is justified from an experimental insight for MnSi$_{1-x}$Ge$_x$~\cite{Fujishiro2019}.

The chemical potential is set to be $\mu=-3.79$, which reproduces a wavenumber of $Q \approx \pi/4$ and a spatial period of $\lambda=2\pi a/\sqrt{3}Q \approx$2.15 nm in agreement with the experimental value of $\lambda=$1.9-2.1 nm for MnSi$_{1-x}$Ge$_x$~\cite{Fujishiro2019}. Here the lattice constant is assumed to be $a=0.465$ nm. In Figs.~\ref{Fig08}, the Fermi surface and its cross section associated with the nesting vectors $\bm Q_1$ and $\bm Q_2$ are shown. 

%%%%%%%%%%%%%%%%%%%%%%%%%%%%%%%%%%%%%%
\begin{figure}[t]
\includegraphics[scale=0.5]{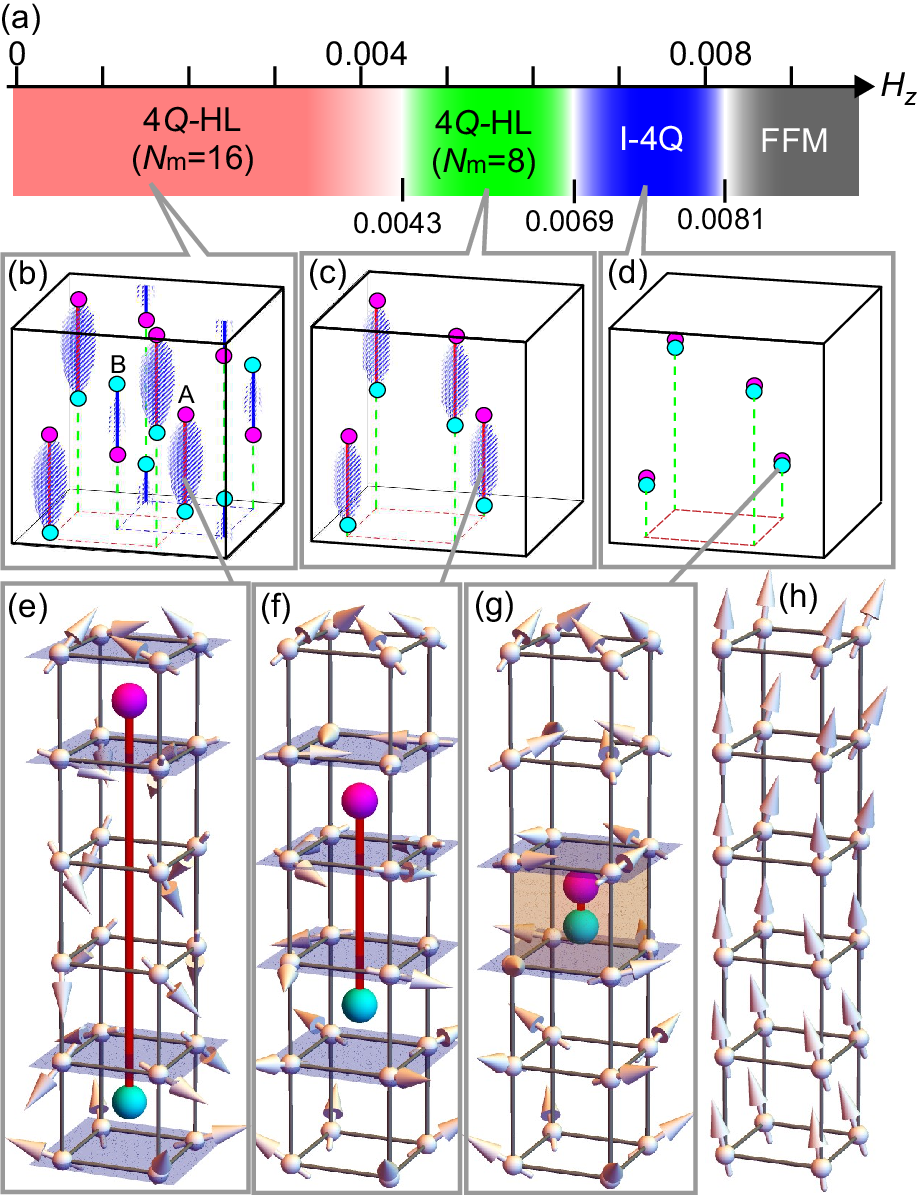}
\caption{(a) Phase diagram of the cubic Kondo-lattice model as a function of $H_z$, which contains two 4$Q$-HL phases, the intermediate 4$Q$ (I-4$Q$) phase, and the forced ferromagnetic (FFM) phase. (b)-(d) Spatial configurations of the Bloch points of hedgehogs (magenta) and antihedgehogs (cyan) in the 4$Q$-HL and I-4$Q$ phases. (e)-(g) Spin configurations around the Dirac string A in respective phases. (h) Spin configurations after the Dirac string A vanishes in the FFM phase.}
\label{Fig09}
\end{figure}
%%%%%%%%%%%%%%%%%%%%%%%%%%%%%%%%%%%%%%
Figure~\ref{Fig09}(a) shows a ground-state phase diagram as a function of magnetic field $H_z$. In the absence of magnetic field ($H_z$=0), a 4$Q$-HL state composed of eight hedgehogs and eight antihedgehogs in the magnetic unit cell appears where the total number of the hedgehogs and antihedgehogs is $N_{\rm m}$=16. This 4$Q$-HL state contains two inequivalent Dirac strings (A and B), each of which connects a hedgehog and an antihedgehog along the $z$ axis. There are eight strings in total, four each for strings A and B. 

The Dirac strings A and B have the same length in the 4$Q$-HL phase with $N_{\rm m}$=16 at $H_z=0$. As $H_z$ increases, both strings A and B gradually shrink. Importantly, the strings B shrink faster than the strings A. Thereby, the strings B disappear first as a result of the hedgehog-antihedgehog pair annihilation when $H_z$ reaches 0.0043, which is a topological phase transition from the 4$Q$-HL with $N_{\rm m}$=16 to another 4$Q$-HL with $N_{\rm m}$=8. Subsequently, the system enters the intermediate 4$Q$ (I-4$Q$) state at $H_z$=0.0069 in which the hedgehog and antihedgehog collide to be merged and share the same cubic unit of the lattice. With further increasing $H_z$, the merging or the pair annihilation of the hedgehog and antihedgehog connected by strings B complete, and the system enters the forced ferromagnetic phase at $H_z$=0.0081.

%%%%%%%%%%%%%%%%%%%%%%%%%%%%%%%%%%%%%%
\begin{figure}[t]
\centering
\includegraphics[scale=0.6]{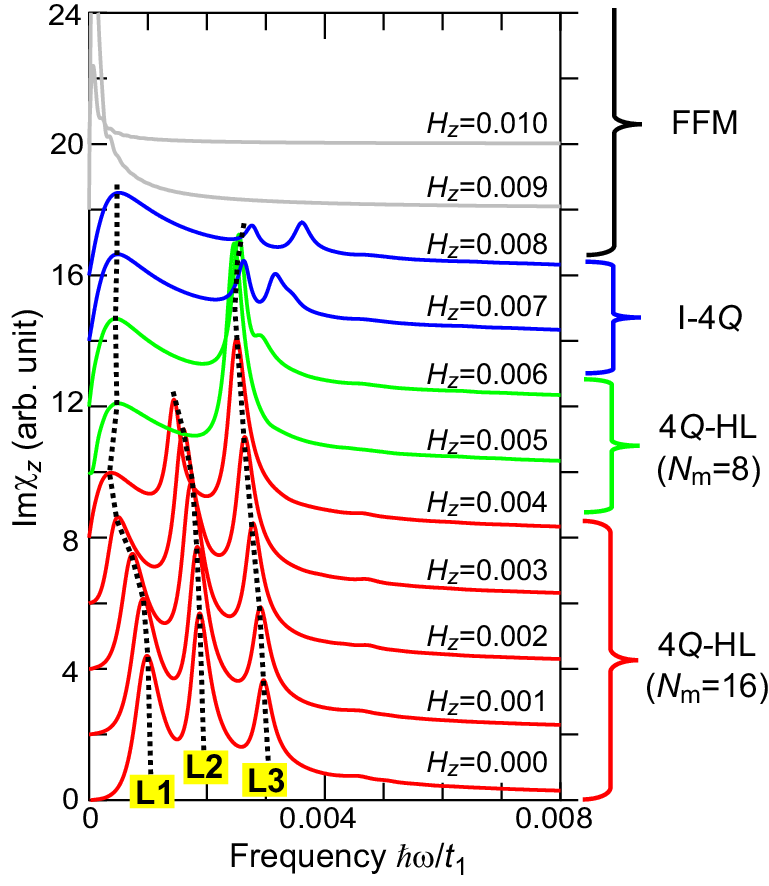}
\caption{Calculated microwave/light absorption spectra for various values of $H_z$.}
\label{Fig10}
\end{figure}
%%%%%%%%%%%%%%%%%%%%%%%%%%%%%%%%%%%%%%
In Fig.~\ref{Fig10}, we show calculated spectra of Im$\chi_z(\omega)$ where $\chi_z(\omega)$ is the longitudinal dynamical magnetic susceptibility. Noticeably, we observe three excitation modes in the 4$Q$-HL phase with $N_{\rm m}=16$, which are named L1, L2, and L3 modes, respectively. These modes are expected to appear in the sub-terahertz regime because $\omega$=0.004 in Fig.~\ref{Fig10} corresponds to 1 THz approximately when we assume $t_1=1$ eV. Calculated spatial profiles of oscillation amplitudes for respective phases inidcate that the L2 mode is associated with Strings B, whereas the L3 mode is with Strings A. Namely, the L2 and L3 modes are localized around the respective Dirac strings. On the contrary, the L1 mode is a nonlocalized mode whose oscillation amplitude is broadly distributed in space.

%%%%%%%%%%%%%%%%%%%%%%%%%%%%%%%%%%%%%%
\begin{figure}[t]
\includegraphics[scale=0.5]{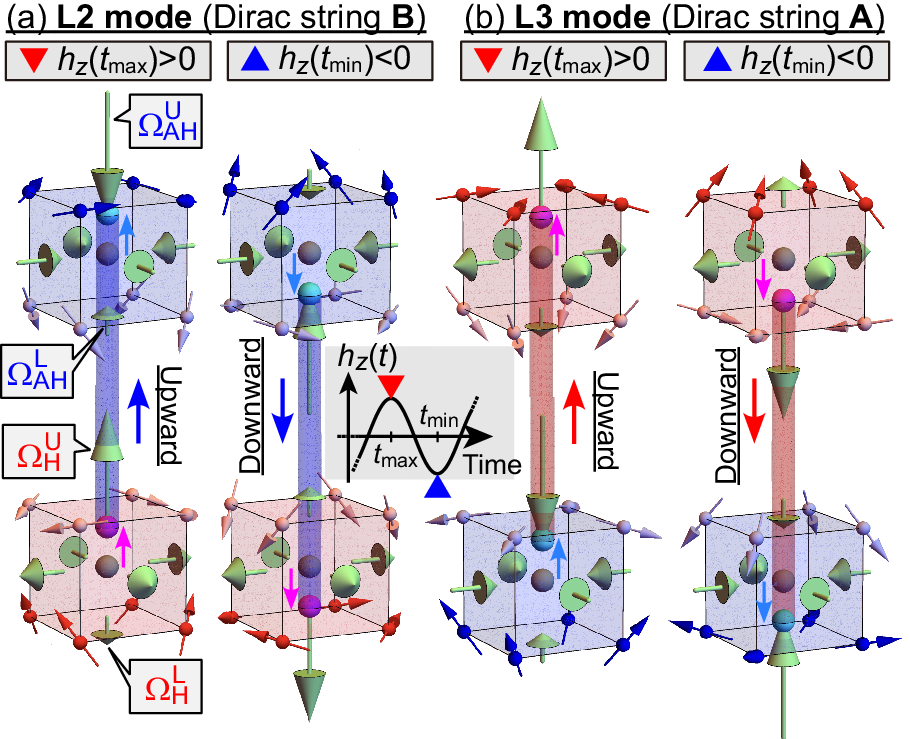}
\caption{Schematics of the translational oscillations of (a) the Dirac string B in the L2 mode and (b) the Dirac string A in the L3 mode.}
\label{Fig11}
\end{figure}
%%%%%%%%%%%%%%%%%%%%%%%%%%%%%%%%%%%%%%
Importantly, the L2 mode disappears when the system enters the 4$Q$-HL phase with $N_{\rm m}=8$ because the Strings B vanish upon this phase transition, which indicates that this topological transition can be detected in the magnetic resonance measurements. On the contrary, the L3 mode survives after this phase transition because the Strings A still exist in the 4$Q$-HL phase with $N_{\rm m}=8$. Moreover, the L3 mode is observed in the $N_{\rm m}=8$ 4$Q$-HL phase and even in the subsequent I-4$Q$ phase in which the merging hedgehog-antihedgehog pair has nonzero scalar spin chirality as a residue of the magnetic topology. The L3 mode disappears when the system enters the forced ferromagnetic phase at $H_z \approx 0.0081$, in which the scalar spin chirality vanishes because of the collinear spin configuration.

The L2 and L3 modes can be regarded as translational modes of Dirac strings B and A, respectively. The ac magnetic field of applied electromagnetic wave dynamically modulates the hedgehog and antihedgehog spin configurations, which results in asymmetric distribution of their emergent magnetic fields on a cubic cell composed of eight spins at its corners. More specifically, the solid angles spanned by four-spin plaquettes on the upper and lower cube faces, which is originally symmetric, becomes asymmetric when the ac magnetic field is applied in the $z$ direction. Thereby, an oscillatory shift of the center-of-mass position of the spin solid angles occurs along the $z$ axis at each hedgehog and antihedgehog. The shifts occur with the same phase in each mode, and the resulting in-phase oscillations can be regarded as coherent translational modes of the Dirac strings [Figs.~\ref{Fig11}(a) and (b)]. It is found that the oscillation phases are the same between L2 and L3 modes, whereas the oscillation phase of the L1 mode is opposite to those of the L2 and L3 modes.

\section{Summary}
In summary, we have overviewed our recent theoretical studies on the spin-charge dynamics and related microwave/light induced dynamical phenomena in the Kondo-lattice models, which exhibit rich long-wavelength magnetic orders including topological magnetism. Several kinds of skyrmion crystals and hedgehog lattices are realized by long-range interactions among the localized spins mediated by itinerant electrons. The long-wavelength magnetic orders in the Kondo-lattice models are usually described by superpositions of multiple helices whose propagation vectors are determined by the nesting of Fermi surfaces. By the large-scale spin-charge dynamics simulations using the Kernel polynomial method, we have revealed a variety of interesting phenomena including (1) microwave-induced switching of magnetic topology for the skyrmion crystal states in the triangular Kondo-lattice model, (2) three linearly dispersive Goldstone modes associated with SO(3) symmetry of the localized spins and one quadratic pseudo-Goldstone mode associated with itinerant-electron charges governed by the discrete translational symmetry of the skyrmion crystal, and unexpected spin-charge segregation in the low-energy excitations of the zero-field skyrmion crystal in the triangular Kondo-lattice model, and (3) two translation modes associated with Dirac strings and disappearance of one of these two modes upon the field-induced topological phase transition with hedgehog-antihedgehog pair annihilations in the quadratic hedgehog lattices in the cubic Kondo-lattice model for real materials of MnSi$_{1-x}$Ge$_x$  and SrFeO$_3$. Recently, new magnetic materials that realize the Kondo-lattice models have been successively discovered and synthesized experimentally. Furthermore, a variety of topological magnetic phases have been observed in these materials. We expect that the phenomena predicted in our studies will be experimentally observed and confirmed in near future. Moreover, we hope that this article could help to open a new research field on fundamental physics and even engineering of rich dynamical magnetic topology in the spin-charge coupled magnets.

 \section{Acknowledgements}
The studies associated with this article have been supported by JSPS KAKENHI (Grants No. 20H00337, No. 23H04522, and No. 24H02231), JST CREST (Grant No. JPMJCR20T1), and Waseda University Grant for Special Research Projects (2023C-140, 2024C-153). R.E. was supported by a Grant-in-Aid for JSPS Fellows (Grant No. 23KJ2047). Most of the numerical calculations were performed using the supercomputers in Institute for Solid State Physics (ISSP), the University of Tokyo.

\end{document}